\makeatletter
\newcommand{\dontusepackage}[2][]{%
  \@namedef{ver@#2.sty}{9999/12/31}%
  \@namedef{opt@#2.sty}{#1}}
\makeatother
\dontusepackage{subfigure}

\documentclass[]{article}

\usepackage{lmodern}
\usepackage{amssymb,amsmath}
\usepackage{ifxetex,ifluatex}
\usepackage[usenames,dvipsnames]{color}
\usepackage{fixltx2e} % provides \textsubscript
\ifnum 0\ifxetex 1\fi\ifluatex 1\fi=0 % if pdftex
  \usepackage[T1]{fontenc}
  \usepackage[utf8]{inputenc}
\else % if luatex or xelatex
  \ifxetex
    \usepackage{mathspec}
    \usepackage{xltxtra,xunicode}
  \else
    \usepackage{fontspec}
  \fi
  \defaultfontfeatures{Mapping=tex-text,Scale=MatchLowercase}
  
\fi
\IfFileExists{upquote.sty}{\usepackage{upquote}}{}
\IfFileExists{microtype.sty}{%
\usepackage{microtype}
\UseMicrotypeSet[protrusion]{basicmath} % disable protrusion for tt fonts
}{}
\usepackage[margin=1.0in,bottom=1.5in]{geometry}
\usepackage[]{natbib}
\usepackage{listings}
\usepackage{graphicx}
\makeatletter
\def\maxwidth{\ifdim\Gin@nat@width>\linewidth\linewidth\else\Gin@nat@width\fi}
\def\maxheight{\ifdim\Gin@nat@height>\textheight\textheight\else\Gin@nat@height\fi}
\makeatother
\setkeys{Gin}{width=\maxwidth,height=\maxheight,keepaspectratio}
\usepackage{caption}
\usepackage{float}
\ifxetex
  \usepackage[setpagesize=false, % page size defined by xetex
              unicode=false, % unicode breaks when used with xetex
              xetex]{hyperref}
\else
  \usepackage[unicode=true]{hyperref}
\fi
\hypersetup{breaklinks=true,
            bookmarks=true,
            pdfauthor={},
            pdftitle={Memory Efficient Invertible Neural Networks for 3D Photoacoustic Imaging},
            colorlinks=true,
            citecolor=black,
            urlcolor=blue,
            linkcolor=black,
            pdfborder={0 0 0}}
\title{Memory Efficient Invertible Neural Networks for 3D Photoacoustic Imaging}
\author{Rafael Orozco\textsuperscript{1}, Mathias Louboutin\textsuperscript{1}
and Felix J. Herrmann\textsuperscript{1}\\Georgia Institute of
Technology\\}
\date{}

\begin{document}
\maketitle
\begin{abstract}
Photoacoustic imaging (PAI) can image high-resolution structures of
clinical interest such as vascularity in cancerous tumor monitoring.
When imaging human subjects, geometric restrictions force limited-view
data retrieval causing imaging artifacts. Iterative physical model based
approaches reduce artifacts but require prohibitively time consuming PDE
solves. Machine learning (ML) has accelerated PAI by combining physical
models and learned networks. However, the depth and overall power of ML
methods is limited by memory intensive training. We propose using
invertible neural networks (INNs) to alleviate memory pressure. We
demonstrate INNs can image 3D photoacoustic volumes in the setting of
limited-view, noisy, and subsampled data. The frugal constant memory
usage of INNs enables us to train an arbitrary depth of learned layers
on a consumer GPU with 16GB RAM.
\end{abstract}

\section{Problem Statement}\label{problem-statement}

In this work, we address the inverse problem of PAI given data $y$. This
data is modeled with a linear forward model $A$ which describes the
propagation of a spatially varying initial acoustic wavefield $x$.
\begin{equation}
y = Ax + \varepsilon \, \, \, \, \text{with} \, \, \, \, \varepsilon \sim \mathcal{N}(0,\sigma^2 I)
\label{eq:problem}
\end{equation}
 We tackle here the realistic case of limited-view, noisy and subsampled
data, which makes the inverse problem ill-posed. Multiple solutions $x$
can explain the data $y$, therefore, this problem is typically solved in
a variational framework where prior knowledge of the solution is
incorporated by solving the minimization of the combination of an
$\ell_2$ data misfit $L$ and a regularization term
$p_{\mathrm{prior}}(x)$. The success of this formulation hinges on users
carefully selecting hyperparameters by hand such as the optimization
step length and the prior itself, typically a multipurpose generic prior
such as TV norm. On the other hand, recent machine learning work argues
that we can learn the optimal step length and even a prior. We pursue
this avenue but select loop-unrolled networks since this learned method
also leverages the known physics of the problem encoded in the model
$A$.

\section{Methods}\label{methods}

\subsection{Loop-unrolled Networks:}\label{loop-unrolled-networks}

\citet{hauptmann2018model} showed that known physical models can be
combined with learned networks by using loop-unrolled networks. These
networks emulate gradient descent where the $i^{\text{th}}$ update is
reformulated as the output of a learned network $\Lambda_{\theta_{i}}$:
\begin{equation}
x_{i+1}, s_{i+1} = \Lambda_{\theta_{i}}(x_{i},s_{i}, \nabla_{x}L(A,x_{i},y))
\label{eq:learnedgd}
\end{equation}
 where $s$ is a memory variable and $\nabla_{x}L$ is the gradient w.r.t
the data misfit on observed data $y$. In our PAI case, $A$ is a linear
operator (discrete wave equation) and $L$ is $\ell_2$ norm data misfit
so the desired gradient is
$\nabla_{x}L(A,x_{i},y) = A^{\top}(Ax_{i}-y)$. Thus, each gradient will
require two PDE solves: one forward and one adjoint. These wave
propagations embed the known physical model into the learned approach.
\citet{hauptmann2018model} show that this method gives promising results
on 3D PAI but note that they are limited to training shallow networks
due to memory constraints. To improve on this method, we propose using
an INN that is an invertible version of loop-unrolled networks.

\subsection{Invertible Neural
Networks:}\label{invertible-neural-networks}

Our invertible neural network (INN) is based on the work of
\citep{putzky2019rim}, where they propose an invertible loop-unrolled
method called invertible Recurrent Inference Machine (i-RIM). While
typical loop-unrolled approaches have linear memory growth in depth
\citet{hauptmann2018model}, i-RIM has constant memory usage due to its
invertible layers. A network constructed with invertible layers does not
need to save intermediate activations thus trading computation cost
required to recompute activations for extremely frugal memory usage. The
memory gains allowed i-RIM authors to train a 480 layer model which was
the state-of-the-art for the FASTMRI challenge when published
\cite{putzky2019rim}. For this work, we adapt i-RIM to Julia and make
our code available alongside other invertible neural networks at
\href{https://github.com/slimgroup/InvertibleNetworks.jl}{InvertibleNetworks.jl}
\citet{witte2020invertiblenetworks}.

\section{Experiments and Results:}\label{experiments-and-results}

\subsection{Generating training
dataset:}\label{generating-training-dataset}

We perform supervised training on ground truth photoacoustic volumes
taken from the blood vessels dataset of \cite{bench2020toward}. For each
3D volume, we simulate photoacoustic data by forward propagating an
initial source at the vessels and restricting the wavefield to a planar
array of receivers at the top of the volume. We then subsample this data
by a factor of 4 and add 10dB SNR Gaussian noise. For wave propagation,
we use Devito a highly optimized DSL \citet{louboutin2019devito}
combined with JUDI \citet{witte2019large}. Propagating a wavefield of
$10^9$ voxels (100x300x300 volume for 1000 time steps) takes 35 seconds
on a 4 core Intel Skylake CPU. We follow the prescription of
\citet{hauptmann2018model} and pre-generate a set of training samples
then fully train one loop-unrolled iteration $\Lambda_{\theta_i}$ at a
time.

\begin{figure}
\centering
\includegraphics[width=1.000\hsize]{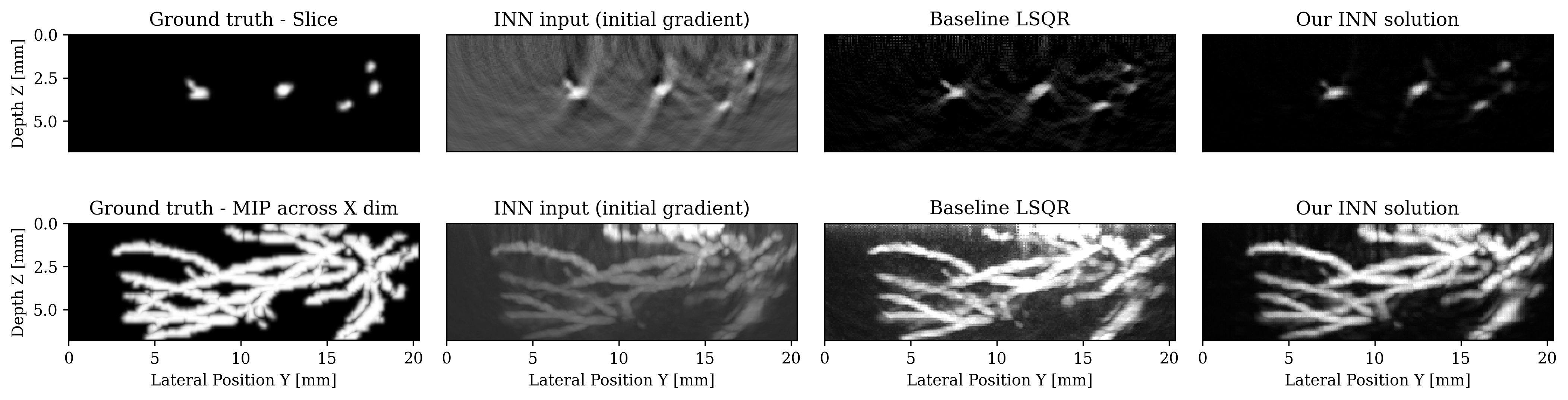}
\caption{First row shows a slice through the 3D volume of: ground truth,
the input to our INN (first gradient w.r.t data misfit on observed data
$y$), baseline LSQR solution after 30 iterations and our final INN
reconstruction after 1 loop-unrolled iteration. Second row shows maximum
intensity projection (MIP) across X dimension.}\label{fig:results}
\end{figure}

\subsection{Results and conclusions:}\label{results-and-conclusions}

After training one loop-unrolled iteration, our INN produces a better
result than the baseline purely physical model based iterative method
(Figure~\ref{fig:results}). For proper comparison, note that our INN
result costs 2 PDE solves ($2 \times 35$ seconds) and a forward pass on
the trained INN (\texttt{7} seconds), while the baseline costs $60$ PDE
solves ($60 \times 35$ seconds). During INN architecture selection, we
observed that the number of layers did not affect memory usage. Thus, we
demonstrated that INNs provide an accurate, memory efficient method that
enables fast 3D imaging for PAI.

\bibliography{midl_2022}

\end{document}